\documentclass{ws-procs9x6}

\begin{document}

\title{Fermionic atoms with tunable interactions in a 3D optical lattice}

\author{T.~St\"oferle, H.~Moritz, C.~Schori, K.~J.~G\"unter, M.~K\"ohl, T.~Esslinger}

\address{Institute of Quantum Electronics, \\
ETH Z\"urich H\"onggerberg, \\
CH--8093 Z\"urich, Switzerland\\
E-mail: stoeferle@phys.ethz.ch}

\maketitle

\abstracts{We report on the realization of a quantum degenerate
atomic Fermi gas in an optical lattice. Fermi surfaces of
noninteracting fermions are studied in a three-dimensional lattice.
Using a Feshbach resonance, we observe a coupling of the Bloch bands
in the strongly interacting regime.}

\section{Introduction}

The exploration of quantum degenerate gases of fermionic atoms is
driven by the ambition to gain deeper insight into long-standing
problems of quantum many-body physics. So far, however, the analogy
to an electron gas in a solid is limited since there the electrons
experience a periodic lattice potential. The lattice structure is in
fact a key ingredient for most models describing quantum many-body
phenomena in materials. We access this regime by preparing a
degenerate atomic Fermi gas in the crystal structure of an optical
lattice.

In our experiment we load a noninteracting gas of $^{40}$K atoms
into a three-dimensional optical lattice with simple cubic symmetry
and directly image the Fermi surfaces. Gradual filling of the
lattice transforms the system from a normal state into a band
insulator. Previous experiments with far-detuned three-dimensional
optical lattices were always carried out with bosonic
atoms\cite{DePue1999,Greiner2002a,Stoferle2004}, and experiments
with fermions were restricted to a single standing
wave\cite{Modugno2003}.

Interactions in the atomic Fermi gas can be tuned by using a
Feshbach resonance between different spin components of the gas. By
ramping the magnetic field into vicinity of the resonance, we
increase the interactions between two particles residing on the same
site of the lattice. In this manner we dynamically induce a coupling
between the lowest energy bands.

\section{Preparing the degenerate Fermi gas in the 3D optical lattice}

The apparatus and the procedure we use to create a degenerate atomic
Fermi gas is described in previous work\cite{Kohl2005b}. In brief,
we use bosonic $^{87}$Rb to sympathetically cool a spin-polarized
gas of fermionic $^{40}$K atoms to temperatures of $T \approx
0.3\,T_\text{F}$ ($T_\text{F}=260\,\text{nK}$ is the Fermi
temperature of the noninteracting gas). The potassium atoms are then
transferred from the magnetic trap into a crossed-beam optical
dipole trap with a wavelength of $\lambda=826\,\text{nm}$. There we
prepare a spin mixture with $(50 \pm 4)\%$ in each of the
$|F=9/2,m_F=-9/2>$ and $|F=9/2,m_F=-7/2>$ spin states using a
sequence of two radio frequency pulses. After further evaporative
cooling in the optical trap, we reach temperatures between
$T=0.2\,T_\text{F}$ and $0.25\,T_\text{F}$ with $5 \times 10^4$ to
$2\times 10^5$ particles, respectively.

Prior to loading the atoms into the optical lattice we tune the
magnetic field to $B=(210 \pm 0.1)\,\text{G}$, such that the
$s$-wave scattering length between the two states vanishes. We
exploit the magnetic Feshbach resonance between the
$|F=9/2,m_F=-9/2>$ and $|F=9/2,m_F=-7/2>$ states\cite{Regal2004b}
which is centered at $B_0=202.1\,\text{G}$ and has a width of
$\Delta B=7.8\,\text{G}$. In this way we produce a two-component
Fermi gas without interactions.

Then the standing wave laser field along the vertical $z$-axis is
turned on. Subsequently, the optical dipole trap along the $y$-axis
is turned off and a standing wave laser field along the same axis is
turned on, followed by the same procedure along the $x$-axis. In
order to keep the loading of the atoms into the lattice as adiabatic
as possible, the intensities of the lasers are slowly increased
(decreased) using exponential ramps with durations of
$20\,\text{ms}$ ($50\,\text{ms}$) and time constants of
$10\,\text{ms}$ ($25\,\text{ms}$), respectively. In its final
configuration the optical lattice is formed by three orthogonal
standing waves with a wavelength of $\lambda=826\,\text{nm}$,
mutually orthogonal polarizations and $1/e^2$-radii of
$50\,\mu\text{m}$ ($x$-axis) and $70\,\mu\text{m}$ ($y$-axis and
$z$-axis). The lattice depth $V_0$ is calibrated by modulating the
laser intensity and studying the parametric heating. The calibration
error is estimated to be $<10\%$.

\section{Observing the Fermi surface}

The potential created by the optical lattice results in a simple
cubic crystal structure. The Gaussian intensity profiles of the
lattice beams give rise to an additional confining potential which
varies with the laser intensity. As a result, the sharp edges
characterizing the $T=0$ distribution function for the quasimomentum
in the homogeneous case are expected to be rounded off.
Noninteracting fermions on a lattice with an additional harmonic
confinement are described by the tight-binding
Hamiltonian\cite{Rigol2004b}
\begin{equation}
    H = -J \sum_{ \langle i,j \rangle} \hat{c}^\dagger_i \hat{c}_j + \frac{m}{2} \sum_i \left( \omega_{\text{ext},x}^2 x_i^2 + \omega_{\text{ext},y}^2 y_i^2 + \omega_{\text{ext},z}^2 z_i^2 \right) \hat{n}_i \,.
    \label{eq:fermions_lattice_noninteracting}
\end{equation}
Here $\hat{c}^\dagger_i$ and $\hat{c}_i$ denote the fermionic
creation and annihilation operators for a particle at site $i$, and
$\hat{n}_i=\hat{c}^\dagger_i \hat{c}_i$ is the number occupation of
site $i$. The first term with the tunneling matrix element $J$
corresponds to the kinetic energy the particles gain through
delocalization over the neighboring lattice sites. The second term
accounts for a site-specific energy offset due to the harmonic
confinement, which is represented by its trapping frequencies
$\omega_{\text{ext},\alpha}$ (with $\alpha=x,y,z$). As the values of
$\omega_\text{ext}$ depend on the intensity of the lattice, changing
the tunnel coupling $J$ by varying the depth of the periodic
potential results in a simultaneous change of the external
confinement.

The harmonic confinement leads to a curvature of the Bloch bands in
position space and, as a consequence, spatial regions with different
fillings. For a given Fermi energy $E_\text{F}$, the filling in the
center is highest and falls off to zero to the outer regions of the
trap. When $E_\text{F}$ exceeds the band width, a band insulating
region appears first in the center of the trap, while regions
farther out still have a filling below unity. However, due to the
curved bands in position space, these atoms may still be spatially
localized to a few lattice sites for small band
widths\cite{Pezze2004}.

In the experiment we probe the population within the Brillouin zones
by ramping down the optical lattice slowly enough for the atoms to
stay adiabatically in the lowest band whilst quasimomentum is
approximately conserved\cite{Greiner2001b}. We lower the lattice
potential to zero over a timescale of $1\,\text{ms}$. After an
additional $1\,\text{ms}$ we abruptly switch off the homogeneous
magnetic field and allow for a total of $9\,\text{ms}$ of ballistic
expansion before we take an absorption image of the expanded atom
cloud. The momentum distribution obtained from these time-of-flight
images reproduces the quasimomentum distributions of the atoms
inside the lattice.

\begin{figure}[ht]
    \centerline{\epsfxsize=\columnwidth\epsfbox{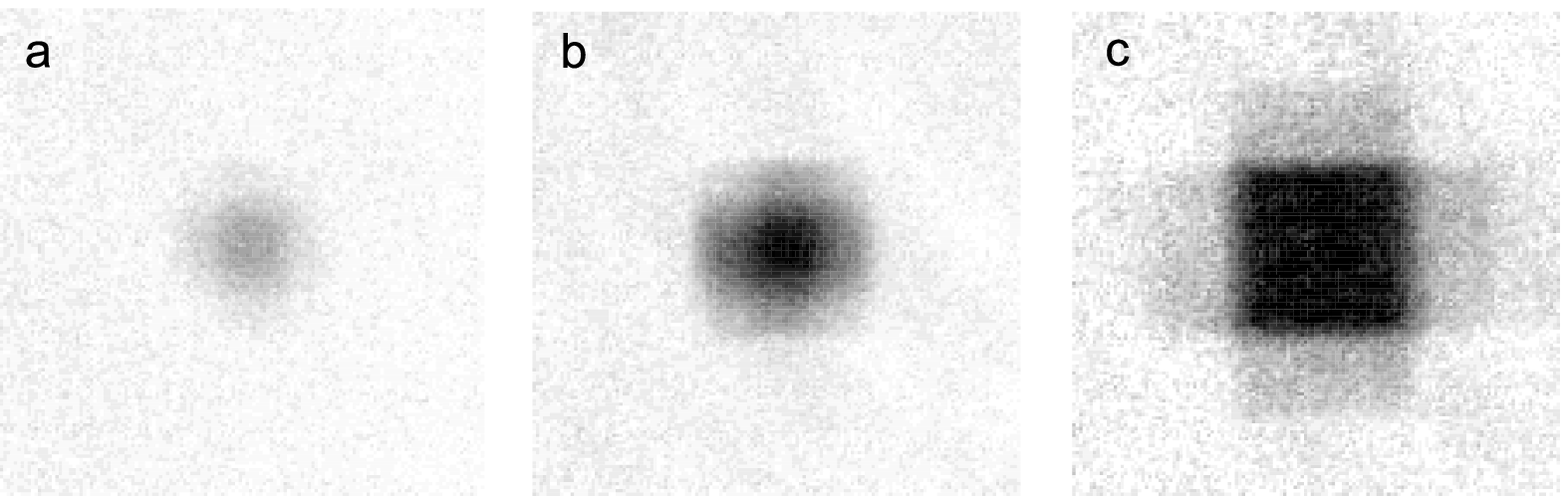}}
    \caption{Development of the Fermi surface in quasimomentum space when more and more
fermions are filled into the lattice. a) 3500 atoms per spin state
and $V_0=5\,E_\text{r}$. b) 15000 atoms per spin state and
$V_0=5\,E_\text{r}$. c) 100000 atoms per spin state and
$V_0=20\,E_\text{r}$. When the Fermi energy exceeds the initial
width of the lowest Bloch band, higher Brillouin zones begin to
become populated. As the band structure of the second and the third
band in a 3D optical lattice overlap partially, both the second and
the third Brillouin zone are being filled with atoms. Each of the
three displayed quasimomentum distributions is obtained by averaging
and smoothing 4 absorption images taken with the same parameters.}
        \label{fig:fermions_lattice_fermisurface_filling}
\end{figure}

With increasing particle number and confinement, the filling of the
lattice sites increases. The initially circular shape of the Fermi
surface (figure~\ref{fig:fermions_lattice_fermisurface_filling}a)
develops extensions pointing towards the Bragg planes
(figure~\ref{fig:fermions_lattice_fermisurface_filling}b) and
finally transforms into a square shape completely filling the first
Brillouin zone deeply in the band insulator. We have observed
population of higher bands if more atoms are filled into the lattice
initially (figure~\ref{fig:fermions_lattice_fermisurface_filling}c).

\section{Tuning the interactions}

By using a Feshbach resonance, it is possible to tune continuously
from attractive to repulsive $s$-wave interactions. When the on-site
interaction is much smaller than the band gap, the physics in the
lowest Bloch band can be described in the tight-binding
approximation by a Hubbard Hamiltonian\cite{Jaksch1998}. The phase
diagram of the Fermi-Hubbard Hamiltonian is very rich, including
Mott insulating\cite{Rigol2003} and antiferromagnetic
phases\cite{Hofstetter2002}. We explore the limit where the wells of
the lattice can be regarded as an array of isolated harmonic
oscillator potentials, and we use a Feshbach resonance to tune the
interactions between the particles in each of the wells.

\begin{figure}[ht]
    \begin{center}
    \centerline{\epsfxsize=\columnwidth\epsfbox{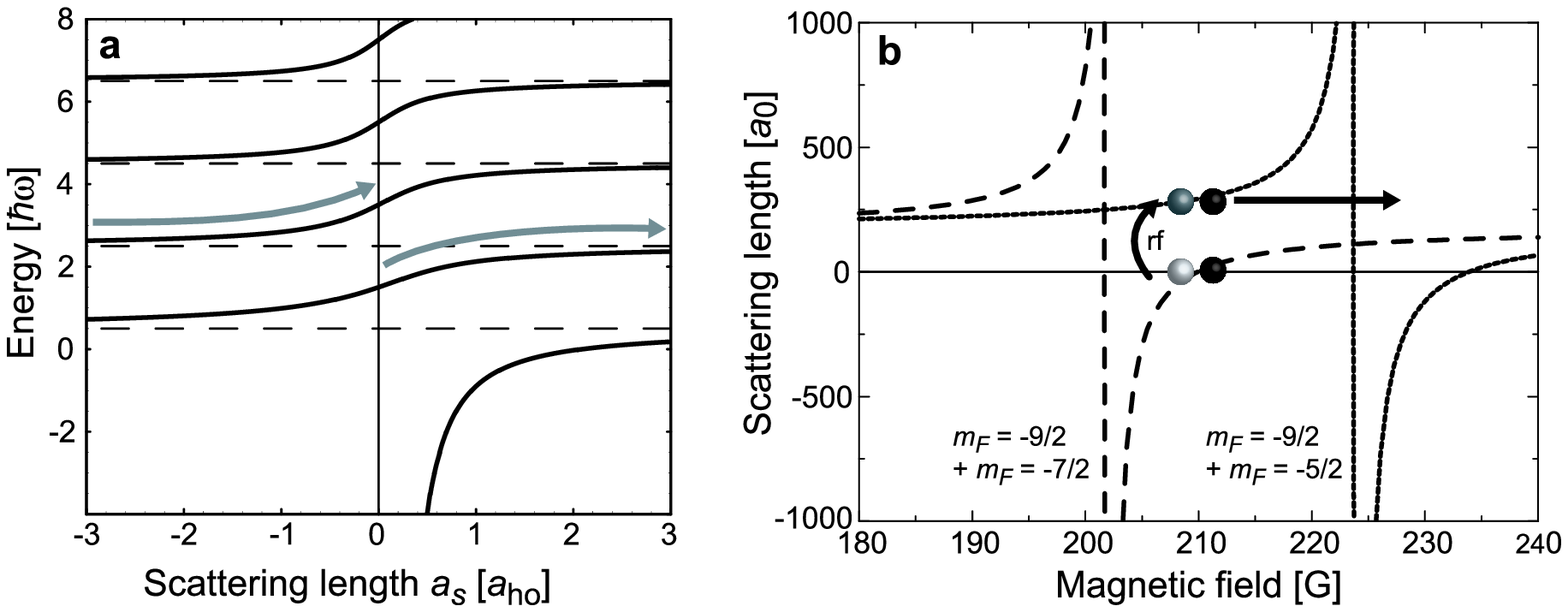}}
        \caption{a) Energy spectrum of two strongly interacting particles in
a three-dimensional harmonic oscillator potential which is given by
equation (\ref{eq:fermions_lattice_interactions_eigenenergies}) in
the center-of-mass frame. The asymptotic energies at $a_s
\rightarrow \pm\infty$ are indicated by the dashed lines. The arrows
show the direction of our sweep over the Feshbach resonance. b)
Experimental method to observe interaction-induced transitions
between Bloch bands. Two Feshbach resonances between the
$|F=9/2,m_F=-9/2>$ and $|F=9/2,m_F=-7/2>$ states (dashed line, left)
and the $|F=9/2,m_F=-9/2>$ and $|F=9/2,m_F=-5/2>$ states (dotted
line, right) are exploited to tune the interactions in the gas.
Using an rf pulse, we transfer the atoms from the $|F=9/2,m_F=-7/2>$
to the $|F=9/2,m_F=-5/2>$ state prior to sweeping over the Feshbach
resonance at $224.2\,\text{G}$. }
        \label{fig:fermions_lattice_interactions_sequence}
    \end{center}
\end{figure}

In deep optical lattice potentials where tunneling is negligible,
the Bloch bands are nearly flat and can be approximated by the
levels of the harmonic oscillator inside each lattice well. When the
$s$-wave interaction is changed on a time scale short compared to
the tunneling time between adjacent potential minima, we may regard
the band insulator as an array of more than ten thousand independent
harmonic potential wells, each of which is populated by two
particles. Solving the Schr\"odinger equation of two particles with
a contact interaction in a harmonic oscillator
yields\cite{Busch1998}
\begin{equation}
    \frac{a_s}{a_\text{ho}} = \left( \sqrt{2} \frac{\Gamma(-E/(2\hbar\omega)+3/4)}{\Gamma(-E/(2\hbar\omega)+1/4)}
    \right)^{-1}
    \label{eq:fermions_lattice_interactions_eigenenergies}
\end{equation}
where $E$ denotes the eigenenergies for a given scattering length
$a_s$, and $\omega$ is the trap frequency. The oscillator length
$a_\text{ho}=\sqrt{\hbar/(m\omega)}$ for a single lattice well in
our system is on the order of $1300\,a_0$.
Figure~\ref{fig:fermions_lattice_interactions_sequence}a displays
the energy spectrum as a function of the scattering length $a_s$.

On the Feshbach resonance, the scattering length has a pole ($a_s
\rightarrow \pm\infty$). When sweeping adiabatically over the
Feshbach resonance from magnetic fields below the resonance, we
first increase the repulsive interaction between the two atoms in
the oscillator ground state from the value of the background
scattering length to $a_s \rightarrow +\infty$ on the resonance.
There the energy level of the lowest harmonic oscillator level
becomes degenerate with the energy of the second next level, which
is lowered from its noninteracting value because of the divergence
$a_s \rightarrow -\infty$ on the resonance. If projected onto the
noninteracting eigenstates, this corresponds to a superposition of
many higher oscillator states. When continuing the magnetic field
sweep to values where the scattering length is again near zero, the
atoms may adiabatically follow the higher oscillator level. This
simple picture of two strongly interacting atoms in a well gives a
qualitative understanding on what happens when atoms in a
three-dimensional optical lattice cross a Feshbach resonance: The
higher energy bands get populated due to interaction induced
coupling between the Bloch bands.

We experimentally investigate the interacting regime in the lattice
starting from a noninteracting gas deep in a band insulator with
$V_x=12\,E_\text{r}$ and $V_y=V_z=18\,E_\text{r}$ and corresponding
trapping frequencies of $\omega_x=2 \pi \times 50\,\text{kHz}$ and
$\omega_y=\omega_z=2 \pi \times 62\,\text{kHz}$ in the individual
minima. A short radio-frequency pulse of $40~\mu\text{s}$ is applied
to transfer all atoms from the $|F=9/2,m_F=-7/2>$ into the
$|F=9/2,m_F=-5/2>$ state, with the atoms in the $|F=9/2,m_F=-9/2>$
remaining unaffected. Then we ramp the magnetic field to different
final values around the Feshbach resonance\cite{Regal2003a} located
at $B_0=224.2\,\text{G}$ which has a width of $\Delta B =
9.7\,\text{G}$. The sweep approaches the Feshbach resonance from the
side of repulsive interactions and crosses the resonant point where
the scattering length diverges towards the side of attractive
interactions, as depicted in
figure~\ref{fig:fermions_lattice_interactions_sequence}b. When using
this direction of the sweep there is no adiabatic conversion to
molecules. After turning off the optical lattice adiabatically and
switching off the magnetic field we measure the momentum
distribution. To see the effect of the interactions we determine the
fraction of atoms transferred into higher bands.

\begin{figure}[ht]
    \centerline{\epsfxsize=\columnwidth\epsfbox{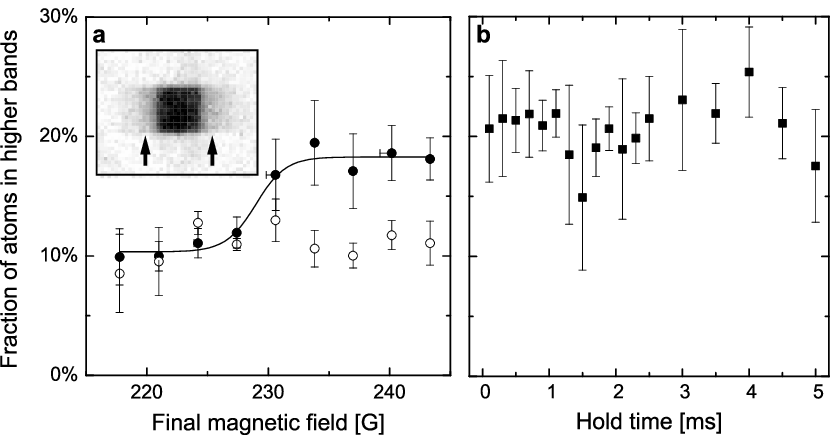}}
    \caption{Interaction induced transition between Bloch bands. a) A
    fraction of the fermions is transferred into higher bands using a sweep across the
Feshbach resonance (filled symbols). The line shows a sigmoidal fit
to the data. The open symbols show a repetition of the experiment
with the atoms prepared in the spin states $|F=9/2,m_F=-9/2>$ and
$|F=9/2,m_F=-7/2>$ where the scattering length is not sensitive to
the magnetic field. The magnetic field is calibrated by rf
spectroscopy between Zeeman levels. Due to the rapid ramp the field
lags behind its asymptotic value and the horizontal error bars
represent this deviation. The data is taken from reference~[1]. The
inset shows the quasimomentum distribution for a final magnetic
field of $233\,\text{G}$, for which the images of $4$ measurements
have been averaged and smoothed. Arrows indicate the atoms in the
higher bands. b) Measured fraction in the higher bands versus the
hold time at $233.8\,\text{G}$ after the sweep. The vertical error
bars show the statistical error of $4$ repetitive measurements.}
        \label{fig:fermions_lattice_interactions_data}
\end{figure}

Figure~\ref{fig:fermions_lattice_interactions_data}a shows the
measurement results when ramping the magnetic field to different
final values around the Feshbach resonance with a fixed inverse
sweep rate of $12\,\mu\text{s}/\text{G}$. For final magnetic field
values well above the Feshbach resonance we observe a significant
increase in the number of atoms in higher bands along the weak axis
of the lattice, demonstrating an interaction induced coupling
between the lowest bands. When we choose symmetrical lattice
intensities, the atoms in the higher bands are distributed over all
three dimensions since the oscillator levels are degenerate. By
using a single, weaker axis we lift this degeneracy and transfer
only to the higher bands along this direction, which increases our
signal-to-noise ratio crucially.

Initially, one might expect that all atoms should be transferred to
the higher bands. However, the process could be limited by the
actual number of doubly occupied lattice sites which constitutes an
upper bound for the transferred fraction. The inhomogeneous filling
of the lattice and the finite temperature give rise to a significant
amount of singly occupied sites, especially in the regions away from
the center. When we vary the hold time at the final magnetic field
far above the resonance, we do not observe a significant dependence
on the measured fraction of atoms in the higher bands
(figure~\ref{fig:fermions_lattice_interactions_data}b). Very recent
theoretical studies\cite{Dickerscheid2005,Diener2005} are beginning to shed light
onto physics beyond the single-band description of the Feshbach
resonance in the lattice.

\section{Conclusions}

In our experiments, we have realized a degenerate Fermi gas with
tunable interactions in a three-dimensional optical lattice. We have
demonstrated the control over parameters of the system such as
filling and interactions. In the noninteracting, static regime, we
image the shape of the Fermi surface for different characteristic
densities. Our measurements in the strongly interacting regime where
we observe a dynamical coupling of the Bloch bands pose challenges
for the present theoretical understanding of many-particle fermionic
systems in optical lattices. This very generic implementation of a
fermionic many-particle quantum system on a lattice is expected to
provide new avenues to intriguing phenomena of quantum many-body
physics. The unique control over all relevant parameters in this
intrinsically pure system allows us to carry out experiments which
are not feasible with solid-state systems.

\section*{Acknowledgments}

We would like to thank SNF and QSIT for funding.


\begin{thebibliography}{0}

\bibitem{Kohl2005b} M. K\"ohl, H. Moritz, T. St\"oferle, K.J. G\"unter, and T. Esslinger, {\it Phys. Rev. Lett.} {\bf 94}, 080403 (2005).
\bibitem{DePue1999} M.~T. DePue, C. McCormick, S.~L. Winoto, S. Oliver, and D.~S. Weiss, {\it Phys. Rev. Lett.} {\bf 82}, 2262 (1999).
\bibitem{Pezze2004} L. Pezz\`e, L. Pitaevskii, A. Smerzi, S. Stringari, G. Modugno, E. de Mirandes, F. Ferlaino, H. Ott, G. Roati, and M. Inguscio, {\it Phys. Rev. Lett.} {\bf 93}, 120401 (2004).
\bibitem{Greiner2002a} M. Greiner, O. Mandel, T. Esslinger, T.~W. H\"ansch, and I. Bloch, {\it Nature} {\bf 415}, 39 (2002).
\bibitem{Stoferle2004} T. St\"oferle, H. Moritz, C. Schori, M. K\"ohl, and T. Esslinger, {\it Phys. Rev. Lett.} {\bf 92}, 130403 (2004).
\bibitem{Modugno2003} G. Modugno, F. Ferlaino, R. Heidemann, G. Roati, and M. Inguscio, {\it Phys. Rev. A} {\bf 68}, 011601(R) (2003).
\bibitem{Regal2004b} C.~A. Regal, M. Greiner, and D.~S. Jin, {\it Phys. Rev. Lett.} {\bf 92}, 083201 (2004).
\bibitem{Rigol2004b} M. Rigol and A. Muramatsu, {\it Phys. Rev. A} {\bf 70}, 043627 (2004).
\bibitem{Greiner2001b} M. Greiner, I. Bloch, O. Mandel, T.~W. H\"ansch, and T. Esslinger, {\it Phys. Rev. Lett.} {\bf 87}, 160405 (2001).
\bibitem{Jaksch1998} D. Jaksch, C. Bruder, J.~I. Cirac, C.~W. Gardiner, and P. Zoller, {\it Phys. Rev. Lett.} {\bf 81}, 3108 (1998).
\bibitem{Rigol2003} M. Rigol, A. Muramatsu, G.~G. Batrouni, and R.~T. Scalettar, {\it Phys. Rev. Lett.} {\bf 91}, 130403 (2003).
\bibitem{Hofstetter2002} W. Hofstetter, J.~I. Cirac, P. Zoller, E. Demler, and M.~D. Lukin, {\it Phys. Rev. Lett.} {\bf 89}, 220407 (2002).
\bibitem{Busch1998} T. Busch, B.-G. Englert, K. Rzazewski, and M. Wilkens, {\it Found. Phys.} {\bf 28}, 549 (1998).
\bibitem{Regal2003a} C.~A. Regal and D.~S. Jin, {\it Phys. Rev. Lett.} {\bf 90}, 230404 (2003).
\bibitem{Dickerscheid2005} D.~B.~M. Dickerscheid, U. Al Khawaja, D. van Oosten, and H.~T.~C. Stoof {\it Phys. Rev. A} {\bf 71}, 043604 (2005).
\bibitem{Diener2005} R.~B. Diener and T.-L. Ho, {\it arXiv e-print}, cond-mat/0507253.

\end{thebibliography}
\end{document}